\documentclass[12pt]{iopart}
\usepackage{amssymb}
\usepackage{graphicx}
\begin{document}

\title[Effect of Coulomb repulsion on the London penetration depth in cuprate]{Effect of Coulomb repulsion on the London penetration depth in cuprate superconductors}

\author{K K Komarov$^{1}$, D M Dzebisashvili$^{1,2}$}

\address{$^{1}$ Kirensky Institute of Physics, Federal Research Center KSC SB RAS 660036, Krasnoyarsk, Russia}
\address{$^{2}$ Reshetnev Siberian State University of Science and Technology 660037, Krasnoyarsk, Russia}
\ead{\mailto{constlike@gmail.com} and \mailto{ddm@iph.krasn.ru}}


\begin{abstract}
We study the effect of Coulomb repulsion between oxygen holes on the London penetration depth $\lambda$ based on the concept of spin-polaron nature of Fermi quasiparticles in cuprates superconductors. It is shown that for the generally accepted values of the parameters of the spin-fermion model, taking into account the Coulomb interaction, both the one-site Hubbard $U_p$ and interaction between holes on the next-nearest-neighbor oxygen ions $V_2$, allows one to achieve a much better agreement of the calculated temperature dependencies of the value $\lambda^{-2}$ with the experimental data in La$_{2-x}$Sr$_{x}$CuO$_{4}$ in a wide range around optimal doping.
\end{abstract}

\vspace{2pc}
\noindent{\it Keywords}: strongly correlated electron systems, Mott-Hubbard materials, high-temperature superconductivity, spin-charge coupling, Coulomb repulsion, London penetration depth

\submitto{Physica Scripta}

%
%
\maketitle
%
%

\section{Introduction}\label{sec:intro}

The existence of strong electron correlations (SEC), due to the significant Coulomb interaction of holes in $d_{x^2-y^2}$-orbitals of copper ions, essentially complicates the study of low-temperature properties of cuprate high-temperature superconductors (HTSC). On the other hand, it is the large value of this interaction that allows to integrate out the high-energy states in the, most realistic for cuprates, three-band $p$$-$$d$ model or the Emery model \cite{Emery_1987,Varma_1987,Hirsch_1987,Gaididei_1988,Ovchinnikov_1989} and to obtain a more simple spin-fermion model (SFM) \cite{Barabanov_1988,Zaanen_1988,Emery_Reiter_1988,Prelovsek_1988,Stechel_1988}. An important difference of the last model from the other effective low-energy models of cuprates, such as the Hubbard model (for example, \cite{Kohno_2018,Kitatani_2019}) or the $t$$-$$J$ model (\cite{Spalek_2017}), is that the SFM clearly takes into account the spatial separation of hole states on the copper ion and two oxygen ions in the unit cell of CuO$_2$-planes.

Within SFM, the concept of a spin polaron was developed \cite{Barabanov_1993,Barabanov_1996,Barabanov_1997}, which made it possible to achieve significant progress in describing the properties of cuprates both in the normal \cite{Barabanov_1997,Starykh_1995,Barabanov_Qpol_1997,Barabanov_2001,Kuzian_2003,DVB_2013}, and superconducting \cite{VDB_PLA_2015,VDB_JLTP_2015,VDB_JSNM_2016} phases. In particular, in \cite{VDB_PLA_2015,VDB_JLTP_2015,VDB_JSNM_2016} it was shown that the Cooper instability develops in an ensemble of spin polarons, and the exchange interaction between the spins localized on copper ions causes an effective attraction between spin-polaron quasiparticles.

Recently, in \cite{Dzebisashvili_2018}, the spin polaron concept was used to describe the dependence of the London penetration depth $\lambda$ on the temperature $T$ in hole-doped cuprate HTSCs. An important result of these studies was the detection of the so-called inflection point in the calculated curves of $\lambda^{-2}(T)$, which was experimentally observed, for example, in La$_{1.83}$Sr$_{0.17}$CuO$_4$ \cite{Khasanov_PRL_2007,Wojek_2011}, YBa$_2$Cu$_3$O$_{7-\delta}$ \cite{Sonier_1999,Khasanov_2007} and Bi$_{2.15}$Sr$_{1.85}$CaCu$_2$O$_{8+\delta}$ \cite{Anukool_2009}.

Unfortunately in \cite{Dzebisashvili_2018} the theoretical curves $\lambda^{-2}(T)$ exceeded the experimental ones for the La$_{2-x}$Sr$_{x}$CuO$_4$ (LSCO) \cite{Panagopoulos_1999} by 30\%-40\%, both regarding the value of $\lambda^{-2}_0$ (i.e. $\lambda^{-2}$ at $T=0$) and the value of $T_c$ which is the temperature at which $\lambda$ diverges. It is important to note that parameters of the SFM were not adjusted, but were chosen equal to those used earlier \cite{Barabanov_Qpol_1997,Barabanov_2001,DVB_2013,VDB_PLA_2015,VDB_JLTP_2015,VDB_JSNM_2016}. To obtain a satisfactory agreement of the $\lambda^{-2}(T)$ curves with the experimental data, it was necessary to reduce by almost two times both the parameter of the spin-fermion coupling $J$, which significantly affects the value of the superconducting current, and the super-exchange parameter $I$, which is the coupling constant in the spin-polaron ensemble, and thus, determining the critical temperature $T_c$. If the two-fold reduction of $J$, used to fit the results in \cite{Dzebisashvili_2018}, could still be somehow justified (the effective parameter $J$ depends on the parameters of the original Emery model and can vary within the specified limits), then the reduction of the exchange integral $I$ was only illustrative.

In this work, it will be shown that taking into account the Coulomb repulsion between the holes on oxygen ions, eliminates the need to artificially underestimate the value of the super-exchange integral to achieve a satisfactory agreement between the theoretical and experimental temperature dependencies of the function $\lambda^{-2}(T)$ in cuprate HTSCs.

The paper is organized as follows. In the second Section, SFM is formulated and necessary notations are introduced. The third Section describes the modification of the SFM Hamiltonian, when the magnetic field is switched on, and the method of calculating the London length. In the fourth Section, the projection method is briefly discussed, on the basis of which the spin polaron concept is implemented, and the system of equations for the Green's functions in the superconducting phase is given. The equations for the order parameter and spectrum of spin-polaron quasiparticles in the superconducting phase are discussed in Section \ref{sec:EqOP}. Section \ref{sec:London} presents the results of numerical calculations of the function $\lambda^{-2}(T)$. The main conclusions of the paper are formulated in the final seventh Section.

\section{Spin-Fermion Model}\label{sec:SFM}

The following ratio between the parameters of the Emery model corresponds to the SEC regime in the cuprate HTSCs:
\begin{equation}\label{cond_1}
\Delta_{pd}\sim(U_d-\Delta_{pd}) \gg t_{pd}>0,
\end{equation}
where $U_d$ is the Coulomb repulsion parameter of two holes on a copper ion, $\Delta_{pd}$ is the charge transfer gap between the hole states on copper and oxygen ions, and $t_{pd}$ is the hybridization parameter between the $d$- and $p$-orbitals on copper and oxygen ions, respectively.

Inequalities (\ref{cond_1}) allow reducing the Emery model and obtaining SFM \cite{Barabanov_1988,Zaanen_1988,Emery_Reiter_1988,Prelovsek_1988,Stechel_1988}. Using the quasi-momentum representation for Fermi operators we write the SFM Hamiltonian in the form \cite{VVV_DDM_etal_2017}
\begin{eqnarray}\label{HamSF}
  \hat{H}_{\mathrm{sp}\textrm{-}\mathrm{f}}=\hat{H}_{\mathrm{h}}+\hat{J}+\hat{I}+\hat{U}_{p}+\hat{V}_{pp},
\end{eqnarray}
where
\begin{eqnarray}
  \hat{H}_{\mathrm{h}}&=
    \sum_{k\alpha}\Bigl(
      \xi_{k_x}a_{k\alpha}^{\dagger}a_{k\alpha}
     +\xi_{k_y}b_{k\alpha}^{\dagger}b_{k\alpha}\nonumber\\
   &\qquad+t_{k}\bigl(a_{k\alpha}^{\dagger}b_{k\alpha}
     +b_{k\alpha}^{\dagger}a_{k\alpha}\bigr)\Bigr),\label{def_Hh}\\
  \hat{J}&=
    \frac{J}{N}\sum_{f, k, q\atop{\alpha, \beta}} e^{if(q-k)}
      u_{k\alpha}^{\dag}\bigl(\vec{S}_f{\vec{\sigma}}_{\alpha\beta}\bigr)u_{q\beta},\label{def_J}\\
  \hat{I}&=
    \frac{I}{2}\sum_{f, \delta}\vec{S}_f\vec{S}_{f+\delta},\label{def_I}\\
  \hat{U}_{p}&=
    \frac{U_p}{N}\sum_{1, 2, 3, 4}\Bigl[
      a^{\dag}_{1\uparrow}a^{\dag}_{2\downarrow}a_{3\downarrow}a_{4\uparrow}
     +(a\to b)\Bigr]~\delta_{1+2-3-4},\label{def_Up}\\
  \hat{V}_{pp}&=
    \frac{4V_1}{N}\sum_{1, 2, 3, 4\atop{\alpha, \beta}}
      \phi_{3-2}~a^{\dag}_{1\alpha}b^{\dag}_{2\beta}b_{3\beta}a_{4\alpha}~\delta_{1+2-3-4}\nonumber\\
    &\quad+\frac{V_2}{N}\sum_{1, 2, 3, 4\atop{\alpha, \beta}}\Bigl[
        \theta^{xy}_{2-3}~a^{\dag}_{1\alpha}a^{\dag}_{2\beta}a_{3\beta}a_{4\alpha} \nonumber\\
      &\quad\qquad\qquad+\theta^{yx}_{2-3}(a\to b)\Bigr]~\delta_{1+2-3-4}.\label{def_V1}
\end{eqnarray}

When writing (\ref{def_Hh}-\ref{def_V1}) the following notations were used
\begin{eqnarray}\label{Definitions}
  &\xi_{k_{x(y)}}=\tilde{\varepsilon}_p+2\tau s_{k,x(y)}^2-\mu,\quad
   \tilde{\varepsilon}_p=\varepsilon_p+2V_{pd},\nonumber\\
  &t_{k}=(2\tau-4t)s_{k,x}s_{k,y},\quad
   s_{k,x(y)}=\sin\bigl(k_{x(y)}/2\bigr),\nonumber\\
  &\phi_{k}=\cos\frac{k_x}{2}\cdot\cos\frac{k_y}{2},\quad
   \theta^{xy(yx)}_k=e^{ik_{x(y)}}+e^{-ik_{y(x)}},\nonumber\\
  &\tau=t_{pd}^2\bigl(1-\eta\bigr)/\Delta_{pd},\quad
   \eta=\Delta_{pd}/\bigl(U_d-\Delta_{pd}-2V_{pd}\bigr),\nonumber\\
  &J=4t_{pd}^2\bigl(1+\eta\bigr)/\Delta_{pd},\quad
   u_{k\beta}=s_{k,x}a_{k\beta}+s_{k,y}b_{k\beta}.
\end{eqnarray}

The $\hat{H}_{\mathrm{h}}$ operator describes holes on oxygen ions. $a_{k\alpha}^{\dagger}(a_{k\alpha})$ denotes the hole creation (annihilation) operators with a quasi-momentum $k$ and with a spin projection $\alpha=\pm1/2$ in the oxygen ion subsystem with the $p_x$-orbitals. Similar operators from the oxygen ion subsystem with the $p_y$-orbitals are denoted by $b_{k\alpha}^{\dagger}(b_{k\alpha})$. The parameter $\varepsilon_p$ corresponds to the bare binding energy of the holes on oxygen ions. This energy is increased by 2$V_{pd}$ taking into account the Coulomb interaction of the oxygen hole with the two nearest copper ions ($V_{pd}$ is the value of this interaction). The integral of the hole hopping between the oxygen ions is denoted by $t$. The parameter $\tau$ is due to hybridization of the $p$- and $d$-orbitals on the copper and oxygen ions. $\mu$ is the chemical potential.

The $\hat{U}_p$ operator defined by (\ref{def_Up}) describes the Hubbard repulsion of two holes on an oxygen ion with the intensity of $U_p$. For brevity, quasi-momenta and spins with the corresponding indices are denoted by numbers, for example: $1\equiv\{k_1, \sigma_1\}$. The Kronecker symbol $\delta_{1+2-3-4}$ accounts for the momentum conservation law: $\delta_{k_1+k_2-k_3-k_4}$. $N$ is the number of unit cells.

\begin{figure}[ht]
\centering
\includegraphics[width=0.4\textwidth]{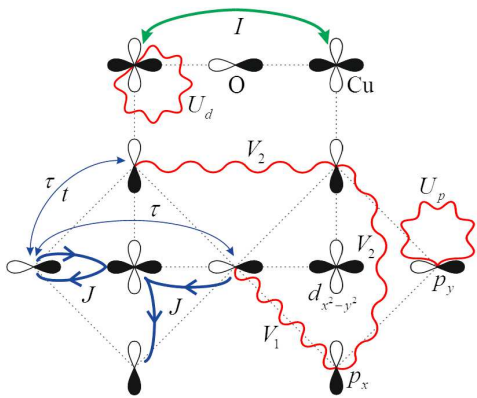}
\caption{(Color online) The structure of the CuO$_2$-plane. Oxygen $p_{x}(p_y)$ orbitals and copper $d_{x^2-y^2}$ orbitals are shown. Wavy lines denote Coulomb interactions: $U_{p(d)}$ --- on-site Coulomb repulsion of holes on an oxygen (copper) ion;  $V_{1}$ and $V_{2}$ --- intersite Coulomb interactions of the holes located at the nearest-neighbor and the next-nearest-neighbor oxygen ions, respectively. The bold green line with arrows stands for the super-exchange interaction ($I$) between spins on the nearest-neighbor copper ions. The bold blue lines next to the letter $J$ correspond to both the spin-fermion exchange interaction and the spin-correlated hoppings. $\tau$ --- effective hole hoppings arising due to $p$$-$$d$-hybridization in the second order of perturbation theory, $t$ --- the integral of direct hole hoppings between nearest oxygen ions (find $\tau$ and $t$ near the thin blue line with arrows).}\label{fig-1}
\end{figure}

Intersite Coulomb interactions of the holes located at the nearest-neighbor and next-nearest-neighbor oxygen ions (figure \ref{fig-1}) are described by the operator $\hat{V}_{pp}$ (see (\ref{def_V1})). The value of these interactions is determined by the parameters $V_1$ and $V_2$, respectively. The functions $\phi_k$ and $\theta^{xy(yx)}_{k}$ appear in the transition from the Wannier representation to the quasi-momentum representation and take into account the crystal symmetry of the CuO$_2$-plane.

The $\hat{J}$ operator appears in the second order in the hybridization parameter $t_{pd}$ and is defined by (\ref{def_J}). This operator takes into account both the exchange interaction between the spins of the holes on copper and oxygen ions, and the spin-correlated hoppings of the hole in the oxygen subsystem with the simultaneous flipping of the localized spin. The spin on the copper ion with the site index $f$ is described by the operator $\vec S_f$, and the vector $\vec{\sigma} $ in (\ref{def_J}) is composed of the Pauli matrices: $\vec{\sigma}=(\sigma^x, \sigma^y, \sigma^z)$.

Finally, the $\hat{I}$ operator takes into account the super-exchange interaction between the nearest-neighbor spins on copper ions and appears in the fourth order of the perturbation theory on the parameter $t_{pd}$. Vector $\delta$ in (\ref{def_I}) connects the site $f$ from the copper sublattice with four nearest sites from the same sublattice.

The SFM parameters --- the effective hopping $\tau$, the integrals of the $p$$-$$d$-exchange ($J$) and super-exchange ($I$) interactions --- are expressed in terms of the parameters of the original Emery model (see, for example, \cite{Zaanen_1988}). The latter are obtained with satisfactory accuracy \cite{Hybertsen_1989,Ogata_2008,Fischer_2011}. Taking this into account as well as the results in \cite{DVB_2013,VDB_JSNM_2016}, we have chosen the following values of the SFM parameters (in eV): $J=1.76$, $I=0.118$, $\tau=0.225$, $U_p=3$ \cite{Hybertsen_1989}. The value of the Coulomb interaction parameters $V_1$ can be estimated in the range $1$$-$$2$ eV \cite{Fischer_2011} albeit, as will be shown below, the particular value of $V_1$ turns out not to be too significant for d-wave superconductivity. The value of $V_2$ we estimated within $0.1$$-$$0.2$ eV according to \cite{VDKB_JMMM_2017}. For the oxygen-oxygen hopping integral we take $t=0.12$ eV which is a reduced value as compared to the one usually used. For choosing this value of $t$ we have at least two reasons following from our previous study of cuprate HTSC in both normal phase \cite{DVB_2013} and d-wave superconducting phase \cite{VDB_JSNM_2016}.

An important circumstance to be taken into account in the spin-polaron approach is that the localized spin subsystem is in the quantum spin-liquid state. This means that the long-range magnetic order is absent in the copper ion subsystem: $\langle S^\alpha_f\rangle=0$ ($\alpha=x, y, z$), but short-range spin correlations remain. These correlations are taken into account through the spin correlation functions $C_j$, which are defined as thermodynamic average of the two spin operators located at a distance $r_j$: $C_j=\bigl\langle{\vec S}_f{\vec S}_{f+r_j}\bigr\rangle$, where $j$ is the number of the coordination sphere of the site $f$. In the spin-liquid phase, these correlators satisfy the sequence of equalities: $C_j=3\bigl\langle S^x_f S^x_{f+r_j}\bigr\rangle=3\bigl\langle S^y_f S^y_{f+r_j}\bigr\rangle=3\bigl\langle S^z_f S^z_{f+r_j}\bigr\rangle.$ In the low temperature range ($\lesssim 100$ K) the spin correlators are almost independent of temperature, but strongly depend on the doping $x$. The correlators $C_j$ as functions of $x$ were calculated, for example, in \cite{Barabanov_2011} based on the frustrated Heisenberg model on a square lattice in the framework of the spherically symmetric approach \cite{Shimahara_1991}. The values of $C_j$ (with $j=1, 2, 3$) used for different $x$ were taken from \cite{Barabanov_2001}.

\section{The London penetration depth}\label{sec:SFMwithA}

Calculation of the penetration depth of the magnetic field $\lambda$ in superconductors is based on the London equation: ${\vec j}=-c/(4\pi\lambda^2) {\vec A}$, where $c$ is the speed of light. In the local approximation this equation establishes a relation between the superconducting current density ${\vec j}$ and the vector potential of the magnetic field ${\vec A}$, and the proportionality coefficient between them is determined by the value of $\lambda$. To calculate the superconducting current density ${\vec j}$ in an ensemble of spin-polaron quasiparticles we should include into Hamiltonian (\ref{HamSF}) terms accounting for coupling to the magnetic field. This can be done via Peierls substitution \cite{Peierls_1933,Lifshitz_1978}. Considering vector potential ${\vec A}_{q}$ in the long-wavelength limit: $q=0$ \cite{Schriffer_1964,Tinkham_1996} we find \cite{Dzebisashvili_2018} that Hamiltonian (\ref{HamSF}) acquires an additional phase
\begin{eqnarray}\label{def_alpfa}
  \alpha_x=\frac{eg_x}{2c\hbar}A^x_{q=0}
\end{eqnarray}
in the argument of the trigonometric function $s_{k, x}$ (\ref{Definitions}). Here $g_x$ is the lattice constant  and for simplicity, we directed the vector potential along the $x$-axis.

Thus, a new definition of the function $s_{k,x}$, which takes into account the magnetic field, has the form:
\begin{eqnarray*}\label{skx_alpha_x}
  s_{k, x}=\sin\bigl(k_x/2-\alpha_x\bigr).
\end{eqnarray*}
It is this definition for $s_{k, x}$ which will be used further. The function $u_{k}$ which is linearly related to $s_{k, x}$ also apparently changes (see (\ref{Definitions})). The function $s_{k, y}$ remains unchanged since in this case $A^y_{q=0}=0$. The Zeeman energy determined by the spin moments of the holes is not taken into account because in the long wavelength limit ($q\to0$) this energy tends to zero.

The resulting expression for the average value of the superconducting current density, obtained in \cite{Dzebisashvili_2018} within SFM, is as follows:
\begin{eqnarray}\label{curr_den}
  j_x(q=0)=\frac{eg_x}{\hbar}\sum_{k\alpha}\cos\Bigl(\frac{k_x}{2}-\alpha_x\Bigr)
   \Bigr[2\tau s_{k, x}\langle a^{\dag}_{k\alpha}a_{k\alpha}\rangle\nonumber\\ \quad
    +\bigl(2\tau-4t\bigr)s_{k,y}\langle a^{\dag}_{k\alpha}b_{k\alpha}\rangle
    +J\langle a^{\dag}_{k\alpha}L_{k\alpha}\rangle\Bigl],
\end{eqnarray}
where expressions for thermodynamic averages in square brackets are given in the Appendix by (\ref{eqCor_abL}). Expression (\ref{curr_den}), in particular, gives the correct behavior of the current density at $T \geq T_c$. Indeed, in the normal phase the dependence of all thermodynamic averages (\ref{eqCor_abL}) on the quasi-momentum $k_x$ is determined only as the difference $k_x-\alpha_x$. Therefore, a simple substitution of the integration variable $k_x \to k_x+\alpha_x$ in the integral in the right part of the expression (\ref{curr_den}) allows one to eliminate the phase $\alpha_x$. Since for $\alpha_x=0$ the integrand in (\ref{curr_den}) is antisymmetric to ${\vec k}$, the right part of (\ref{curr_den}), as required, vanishes.

In the superconducting phase (for $T<T_c$), the dependence of the thermodynamic averages on $k_x$ is determined both by the difference $k_x-\alpha_x$ and by the sum of $k_x+\alpha_x$. In this case, the integral in (\ref{curr_den}) is nonzero.

The inverse square of penetration depth $\lambda^{-2}$ was determined numerically according to the London equation at $T<T_c$ as
\begin{eqnarray*}
  \frac{1}{\lambda^2}=-\frac{4\pi}{c}\cdot\frac{j_x(q=0)}{A^x_{q=0}},
\end{eqnarray*}
where supercurrent density $j_x(q=0)$ is defined by (\ref{curr_den}).

The described approach for calculating $\lambda$ is a sufficiently effective one, especially for multi-band systems, for which the analytical dependence of the quasiparticle spectrum on the quasi-momentum is unknown and can only be obtained numerically. The proposed approach is also convenient since there is no need to carry out cumbersome calculations connected with extracting paramagnetic and diamagnetic parts of the supercurrent density.

\section{Equations for Green's Functions}\label{sec:equations}

A significant feature of the Hamiltonian of the SFM (\ref{HamSF}) is a large value of the $p$$-$$d$-exchange interaction constant $J$, which greatly exceeds the values of all the other parameters of the model. This means that in calculating the energy structure of spin-polaron excitations and analyzing the conditions for superconducting pairing, one has to take into account this interaction exactly. An approach taking into consideration this strong $p$$-$$d$-exchange coupling and within which the corresponding spin-polaron quasiparticle appears is called the spin-polaron approach. For the particular implementation of this approach the Zwanzig-Mori projection technique has proved to be rather convenient \cite{Zwanzig_1961,Mori_1965,Roth_1968,Roth_1969,Rowe_1968,Tserkovnikov_1981,Plakida_book_2010,Mancini_2004}.

According to the projection technique, first of all, it is necessary to introduce a minimal set of basis operators that allow one to correctly describe the quasiparticle excitations in the system. For the correct account of the strong spin-charge coupling in the SFM of interest, it is important to introduce into the specified basis, along with the bare hole operators $a_{k\alpha}$ and $b_{k\alpha}$, the operator
\begin{equation*} \label{L_operator}
  L_{k\alpha}=\frac1N\sum_{fq\beta} e^{if(q-k)}
    \bigl(\vec{S}_f\vec{\sigma}_{\alpha\beta}\bigr)u_{q\beta},
\end{equation*}
arising in the right part of the equations of motion for $a_{k\alpha}$ and $b_{k\alpha}$. As was shown in \cite{Barabanov_1993,Barabanov_1996,Barabanov_1997,Barabanov_2001} the three operators $a_{k\alpha}$, $b_{k\alpha}$ and $L_{k\alpha}$ are sufficient to describe spectral properties of Fermi excitations of the cuprate HTSCs in the normal phase. To analyze the conditions for Cooper instability the mentioned set of three operators, is necessary to be enlarged by three extra operators: $a_{-k\bar{\alpha}}^{\dag}$, $b_{-k\bar{\alpha}}^{\dag}$, $L_{-k\bar{\alpha}}^{\dag}$ ($\bar{\alpha}=-\alpha$) \cite{VDB_PLA_2015,VDB_JLTP_2015,VDB_JSNM_2016}, giving an opportunity to introduce anomalous thermodynamic averages.

The next step of the projection technique is to project the equations of motion for the basis operators (or for the corresponding Green's functions) on the original set of basis operators. The application of this method to the SFM (\ref{HamSF}) with the above basis of six operators is described in \cite{Barabanov_2001,VDB_PLA_2015,VVV_DDM_etal_2017}. Omitting the details of the calculations, we give the answer for a closed system of equations for the Green's functions ($j=1, 2, 3$):
\begin{eqnarray}\label{EqM_GF}
(\omega-\xi_{x})G_{1j}&=\delta_{1j}+t_{k}G_{2j}+J_{x}G_{3j}
  +\Delta_{1k}F_{1j}+\Delta_{2k}F_{2j},\nonumber\\
(\omega-\xi_{y})G_{2j}&=\delta_{2j}+t_{k}G_{1j}+J_{y}G_{3j}
  +\Delta_{3k}F_{1j}+\Delta_{4k}F_{1j},\nonumber\\
(\omega-\xi_{L})G_{3j}&=\delta_{3j}K_{k}+\bigl(J_{x}G_{1j}
  +J_{y}G_{2j}\bigr)K_{k}+\frac{\Delta_{5k}}{K_k}F_{3j},\nonumber\\
(\omega+\xi_{x})F_{1j}&=\Delta_{1k}^*G_{1j}
  +\Delta_{3k}^*G_{2j}-t_{k}F_{2j}+J_{x}F_{3j},\nonumber\\
(\omega+\xi_{y})F_{2j}&=\Delta_{2k}^*G_{1j}
  +\Delta_{4k}^*G_{2j}-t_{k}F_{1j}+J_{y}F_{3j},\nonumber\\
(\omega+\xi_{L})F_{3j}&=\frac{\Delta^*_{5k}}{K_k}G_{3j}
  +\bigl(J_{x}F_{1j}+J_{y}F_{2j}\bigr)K_{k}.
\end{eqnarray}
Here, for the normal and anomalous Green's functions, we use the short notations $G_{ij}$ and $F_{ij}$, respectively. The meaning of these designations is revealed by the equalities:
\begin{eqnarray*}
&G_{11}=\bigl\langle\bigl\langle a_{k\uparrow}|
  a_{k\uparrow}^{\dag}\bigr\rangle\bigr\rangle_\omega,
&F_{11}=\bigl\langle\bigl\langle a_{-k\downarrow}^{\dag}|
  a_{k\uparrow}^{\dag}\bigr\rangle\bigr\rangle_\omega,\nonumber\\
&G_{21}=\bigl\langle\bigl\langle b_{k\uparrow}|
  a_{k\uparrow}^{\dag}\bigr\rangle\bigr\rangle_\omega,
&F_{21}=\bigl\langle\bigl\langle b_{-k\downarrow}^{\dag}|
  a_{k\uparrow}^{\dag} \bigr\rangle\bigr\rangle_\omega,\nonumber\\
&G_{31}=\bigl\langle\bigl\langle L_{k\uparrow}|
  a_{k\uparrow}^{\dag}\bigr\rangle\bigr\rangle_\omega,\quad
&F_{31}=\bigl\langle\bigl\langle L_{-k\downarrow}^{\dag}|
  a_{k\uparrow}^{\dag}\bigr\rangle\bigr\rangle_\omega.
\end{eqnarray*}
The functions $G_{i2}$($F_{i2}$) and $G_{i3}$($F_{i3}$) ($i=1, 2, 3$) are defined in a similar way except for the operator $a^{\dag}_{k\uparrow}$ being substituted for $b^{\dag}_{k\uparrow}$ and $L^{\dag}_{k\uparrow}$, respectively. When writing the system (\ref{EqM_GF}) we use the functions:
\begin{eqnarray}\label{notations}
\xi_{x(y)}=&\xi_{k_{x(y)}},\quad J_{x(y)}=Js_{k, x(y)},\nonumber\\
\xi_L(k)=&\tilde\varepsilon_p-\mu-2t+5\tau/2-J-\tau C_1\gamma_{1k}/2 \nonumber\\
&+\bigl[(\tau-2t)\bigl(-C_1\gamma_{1k}+C_2\gamma_{2k}\bigr)
  +\tau C_3\gamma_{3k}/2\nonumber\\
&+JC_1(1+4\gamma_{1k})/4-IC_1(\gamma_{1k}+4)\bigr]/K_{k},
\end{eqnarray}
where
\begin{eqnarray}
  K_{k}=\bigl\langle\{L_{k\uparrow}, L^{\dag}_{k\uparrow}\}\bigr\rangle=3/4-C_1\gamma_{1k},
\end{eqnarray}
and $\gamma_{jk}$ ($j=1, 2, 3$) denote the square lattice invariants:
\begin{eqnarray}
\gamma_{1k}&=(\cos(k_x-2\alpha_x)+\cos k_y)/2,\nonumber\\
\gamma_{2k}&=\cos (k_x-2\alpha_x)\,\cos k_y,\nonumber\\
\gamma_{3k}&=(\cos(2k_x-4\alpha_x)+\cos 2k_y)/2,
\end{eqnarray}
taking into account the magnetic field through the phase $\alpha_x$.

The components of the superconducting order parameter $\Delta_{jk}$ are defined as anomalous thermodynamic averages:
\begin{eqnarray}\label{def_Deltas}
\Delta_{1k}&=\bigl\langle\bigl\{\bigl[a_{k\uparrow}, \hat
  H_{\mathrm{sp}\textrm{-}\mathrm{f}}\bigr],
  a_{-k\downarrow}\bigr\}\bigr\rangle,\nonumber\\
\Delta_{2k}&=\bigl\langle\bigl\{\bigl[a_{k\uparrow}, \hat
  H_{\mathrm{sp}\textrm{-}\mathrm{f}}\bigr],
  b_{-k\downarrow}\bigr\}\bigr\rangle,\nonumber\\
\Delta_{3k}&=\bigl\langle\bigl\{\bigl[b_{k\uparrow}, \hat
  H_{\mathrm{sp}\textrm{-}\mathrm{f}}\bigr],
  a_{-k\downarrow}\bigr\}\bigr\rangle,\nonumber\\
\Delta_{4k}&=\bigl\langle\bigl\{\bigl[b_{k\uparrow}, \hat
  H_{\mathrm{sp}\textrm{-}\mathrm{f}}\bigr],
  b_{-k\downarrow}\bigr\}\bigr\rangle,\nonumber\\
\Delta_{5k}&=\bigl\langle\bigl\{\bigl[L_{k\uparrow}, \hat
  H_{\mathrm{sp}\textrm{-}\mathrm{f}}\bigr],
  L_{-k\downarrow}\bigr\}\bigr\rangle.
\end{eqnarray}

\section{Equations for the superconducting order parameters and spin-polaron spectrum}\label{sec:EqOP}

The equations for the components of the superconducting order parameter $\Delta_{jk}$ ($j=1, \dots, 5$) are obtained after calculating the commutators (and anticommutators) in the right hand part of formulas (\ref{def_Deltas}) and projecting the result of the calculations on the introduced basis of six operators. Since, according to the results of \cite{VVV_DDM_etal_2017}, the s-wave superconductivity in the SFM does not occur, when writing equations for $\Delta_{jk}$, we keep only those terms which correspond to the d-wave pairing. The result is given in the appendix by (\ref{Deltas}). The components $\Delta_{2k}$ and $\Delta_{3k}$ for the d-wave pairing turn out to be zero. It is important to note that in the expressions (\ref{Deltas}) for $\Delta_{jk}$ the Coulomb repulsion parameter between the holes located on the nearest-neighbor oxygen ions $V_1$ is missing, since according to \cite{Izyumov_1999,VDKB_2016} it should not contribute to the d-wave pairing.

Anomalous thermodynamic averages in the system of equations (\ref{Deltas}) are calculated using the spectral theorem \cite{Zubarev_1960} and corresponding Green's functions of the system (\ref{EqM_GF}). To analyze the conditions for Cooper instability, it is sufficient to calculate the anomalous averages in the linear approximation with respect to the components $\Delta_{jk}$. As a result, a closed set of homogeneous integral equations for the components of the superconducting order parameter $\Delta^*_{lk}$ ($l=1, 4, 5$) is obtained as follows
\begin{eqnarray}\label{Deltas_spectral_theorem}
\Delta_{1k}^*=&-\bigl(\cos k_x-\cos k_y\bigr)\frac{2V_2}{N}\sum_{lq}
  \cos q_xM^{(l)}_{11}(q)\Delta_{lq}^*,\nonumber\\
\Delta_{4k}^*=&-\bigl(\cos k_x-\cos k_y\bigr)\frac{2V_2}{N}\sum_{lq}
  \cos q_xM^{(l)}_{22}(q)\Delta_{lq}^*,\nonumber\\
\Delta_{5k}^*=&+\bigl(\cos k_x-\cos k_y\bigr)\frac{I}{N}\sum_{lq}
  \bigl(\cos q_x-\cos q_y\bigr)\nonumber\\
&\qquad\times\Bigl(M^{(l)}_{33}(q)-C_1M^{(l)}_{uu}(q)\Bigr)\Delta_{lq}^*\nonumber\\
&+\frac{U_p}{N} \sum_{lq}C_1\Bigl(\cos(k_x-2\alpha_x) M^{(l)}_{11}(q)\nonumber\\
&\quad\qquad\qquad+\cos k_y M^{(l)}_{22}(q)\Bigr)\Delta_{lq}^*\nonumber\\
&-\bigl(\cos k_x-\cos k_y\bigr)\frac{2V_2}{N}\sum_{lq}C_1\cos q_x\nonumber\\
&\qquad\times\Bigl(M^{(l)}_{11}(q)+M^{(l)}_{22}(q)\Bigr)\Delta_{lq}^*.
\end{eqnarray}
When writing (\ref{Deltas_spectral_theorem}) we introduced the following functions
\begin{eqnarray}\label{eqM}
M^{(l)}_{uu}(q)=&-s_{q,x}^2M^{(l)}_{11}(q)-s_{q,y}^2M^{(l)}_{22}(q)\nonumber\\
&-s_{q,x}s_{q,y}\bigl(M^{(l)}_{12}(q)+M^{(l)}_{21}(q)\bigr),\\\label{eqMnm}
M^{(l)}_{nm}(q)=&\sum_{j=1,4}\frac{f(-E_{jq})}{2(-1)^{j+1}E_q(E_{jq}
 -\epsilon_{2q})(E_{jq}-\epsilon_{3q})}\nonumber\\
&\times\frac{S^{(l)}_{nm}(q,E_{jq})}{(E_{jq}+\epsilon_{2,-q})(E_{jq}+\epsilon_{2,-q})},
\end{eqnarray}
where $f(E)=1/(\exp\{E/T\}+1)$ is the Fermi-Dirac distribution function, $\epsilon_{jk}$ and $E_{jk}$ are the energies of quasiparticles in the normal and superconducting states, respectively and $S_{ij}^{(l)}(k, \omega)$ are functions defined in the Appendix (\ref{eqsSij}).

The spectrum of Fermi excitations in the normal phase consists of three branches $\epsilon_{jk}$ ($j=1, 2, 3$) and is determined from the solution of the third order dispersion equation
\begin{eqnarray}\label{det}
\mathrm{det}_{k}(\omega)=&+\bigl(\omega-\xi_{x}\bigr)
\bigl(\omega-\xi_{y}\bigr)\bigl(\omega-\xi_{L}\bigr)\nonumber\\
&-2J_{x}J_{y}t_{k}K_{k}-\bigl(\omega-\xi_{y}\bigr)J_{x}^2K_{k}\nonumber\\
&-\bigl(\omega-\xi_{x}\bigr)J_{y}^2K_{k}-\bigl(\omega-\xi_{L}\bigr)t_{k}^2=0,~
\end{eqnarray}
following from condition of existence of nontrivial solution of the system (\ref{EqM_GF}) at $\Delta_{jk}=0$. With the doping levels $x$ typical for cuprates, the dynamics of the holes on oxygen ions is determined solely by the lower band with the dispersion $\epsilon_{1k}$. This branch of the spectrum is characterized by a minimum in the vicinity of ($\pi/2, \pi/2$) point of the Brillouin zone and is significantly separated from the two upper branches $\epsilon_{2k}$ and $\epsilon_{3k}$. The appearance of the lower branch is due to the strong spin-charge coupling, which induces an exchange interaction between the holes and localized spins at the nearest copper ions, as well as spin-correlated hoppings. The features of the spectrum $\epsilon_{1k}$ without magnetic field were discussed in \cite{VDB_PLA_2015}. In our case, taking into account the magnetic field is of fundamental importance.

Since the chemical potential $\mu$ in the systems under consideration lies in the lower band with the dispersion $\epsilon_{1k}$, and the upper bands, as was mentioned above, are separated by a large energy gap, the spectra $\epsilon_{2k}$ and $\epsilon_{3k}$ are almost unchanged with transition to the superconducting phase: i.e. $E_{jk}=\epsilon_{jk}$ for $j=2, 3$. Obtaining an expression for the spectrum $E_{1k}$ for the lower spin-polaron band in the superconducting phase and in the weak magnetic field is described in detail in \cite{VDKKB_2019}. The expression for the spectrum $E_{1k}$ has the form
\begin{eqnarray}\label{Ek}
E_{1k}=\delta\epsilon_{1k}+ \sqrt{\epsilon^2_{1k}+\Delta^2_k},
\end{eqnarray}
where $\delta\epsilon_{1k}$ is a correction to the polaron spectrum in the normal phase $\epsilon_{1k}$, which is antisymmetric in $k$ and linear in $\alpha_x$, and the gap function $\Delta^2_k$ is expressed as a sum of squares of the components of the superconducting order parameter
\begin{eqnarray}\label{Delta2_k}
\Delta^2_k=|\Delta_{1k}|^2+|\Delta_{4k}|^2+|\Delta_{5k}|^2/K^2_k.
\end{eqnarray}

Note that formally, in the sum over $j$ in the right hand side of expression (\ref{eqMnm}) it is necessary to take into account all the bands. However, since the upper bands (with $j=2, 3$) are empty, their contributions can be ignored. The value of the index $j=4$ in the sum over $j$ in (\ref{eqMnm}) corresponds to the spectrum $E_{4k}=-E_{1, -k}$.

One can see from the system of equations (\ref{Deltas_spectral_theorem}) that the kernels of the integral equations are split, and the solutions of this system are to be found in the following form
\begin{eqnarray}\label{eqs_solution}
\Delta_{1k}=&B_{11}\bigl(\cos k_x-\cos k_y\bigr),\nonumber\\
\Delta_{4k}=&B_{41}\bigl(\cos k_x-\cos k_y\bigr),\nonumber\\
\Delta_{5k}=&B_{51}\cos k_x+B_{52}\cos k_y+B_{53}\bigl(\cos k_x-\cos k_y\bigr)\nonumber\\
&+B_{54}\bigl(\cos k_x-\cos k_y\bigr),
\end{eqnarray}
where the six amplitudes $B_{ij}$ determine the contribution of the corresponding basis functions to the expansion of the order parameter components.

Substituting expansion (\ref{eqs_solution}) into equations (\ref{Deltas_spectral_theorem}) and equating the factors of the corresponding trigonometric functions, we obtain a system of six algebraic equations for determining the amplitudes $B_{ij}$. It is also necessary to add to this system an equation for self-consistently finding the chemical potential $\mu$:
\begin{eqnarray}\label{Eq_mu}
x=&\frac2N\sum_k\sum_{j=1,4} \frac{f(E_{jk})}
  {(-1)^{j+1}2E_{k}(E_{jk}-\varepsilon_{2k})(E_{jk}-\varepsilon_{3k})}\nonumber\\
&\times\frac{R^{x}(k,E_{jk})}{(E_{jk}+\varepsilon_{2,-k})(E_{jk}+\varepsilon_{3,-k})},
\end{eqnarray}
where the function $R^{x}(k,\omega)$ is given in (\ref{eqqRx}).

Numerical calculations show that the following relations between the amplitudes hold: $B_{11}=B_{41}\thickapprox-B_{51}=B_{52}$, $B_{54}/B_{51}\approx-10$, $B_{54}/B_{53}\approx-10^{2}$. Thus, it is seen that the largest contribution to the order parameter component $\Delta_{5k}$ gives the amplitude $B_{54}$, proportional to the exchange integral $I$. Regarding this exchange integral, it should be noted that its value depends on the doping $x$. In \cite{Barabanov_2001}, when calculating the exchange integral in the framework of the Heisenberg model, the effect of doping was simulated by the frustration of the exchange couplings. In accordance with \cite{Barabanov_2001}, we used the product $I(1-p)$ as the exchange integral, where $p$ is the frustration parameter varying from 0.15 to 0.275 with $x$ increasing from 0.03 to 0.22.

\section{Results and discussion}\label{sec:London}

Calculations of the temperature dependence of the magnetic penetration depth $\lambda$ taking into account the one-site Hubbard repulsion of holes and the Coulomb interaction between holes on the next-nearest-neighbor oxygen ions were carried out numerically based on expression (\ref{curr_den}) and self-consistent solution of the system of algebraic equations for the amplitudes $B_{ij}$ together with chemical potential equation (\ref{Eq_mu}).  It is important to note that, except for $t$, the rest of the Emery model parameters were chosen to be equal to those that are generally accepted for hole-doped cuprate HTSCs.
\begin{figure}
\begin{center}
\includegraphics[width=0.4\textwidth]{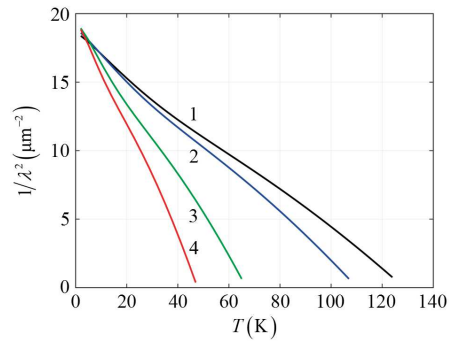}
\caption{The effect of Coulomb repulsion on the temperature dependence of inverse square of the London penetration depth in the SFM of cuprate HTSCs. Curve 1 is calculated with the value of the Coulomb interaction parameters $U_p=V_2=0$; curve 2 --- for $U_p=0$, $V_2=0.1$ eV; curve 3 --- for $U_p=3$ eV, $V_2=0$; curve 4 --- for $U_p=3$ eV, $V_2=0.1$ eV.The value of $V_1$ is not specified since according to (\ref{Deltas_spectral_theorem}) it does not contribute to the d-wave pairing in the SFM. The other model parameter are (in eV): $\tau=0.225$, $t=0.12$, $J=1.76$, $I=0.12$ and $\alpha_x=0.002$, $x=0.17$.}\label{fig-2}
\end{center}
\end{figure}

The calculation results are presented in figure \ref{fig-2}. Curve 1 in this figure is given for comparison. It shows the dependence $\lambda^{-2}(T)$ in the absence of Coulomb interactions ($U_p=V_1=V_2=0$). The remaining curves demonstrate modification of the temperature dependence of $\lambda^{-2}$ with successive switching on the interactions. Note that the parameter $V_1$, as was said above, does not enter the set of equations for order parameter (\ref{Deltas_spectral_theorem}) \cite{VDKB_2016} and, therefore, does not affect the function $\lambda^{-2}(T)$. Curve 2 is obtained by take into account the interactions between the second neighbors; curve 3 --- only the Hubbard repulsion; and curve 4 --- both types of interaction. It is seen that the effect of the Coulomb interaction, in full agreement with the results of \cite{VVV_DDM_etal_2017,VDKB_JLTP_2018}, is manifested in a significant decrease in the critical temperature of the transition to the superconducting phase. The resulting decrease in $T_c$ allows us to achieve a much better agreement of the calculated temperature dependencies of $\lambda^{-2}$ with the experimental data.

\begin{figure}[t]
\begin{center}
\includegraphics[width=0.4\textwidth]{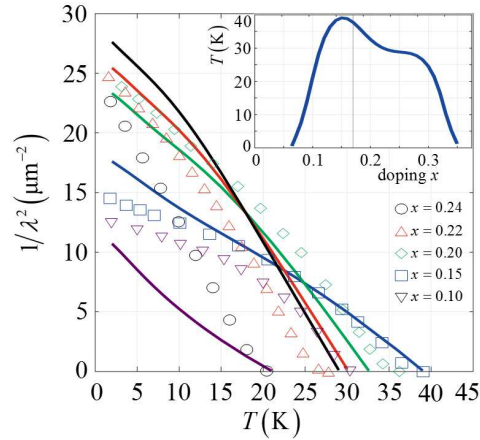}
\caption{(Color online) Temperature dependence of the inverse square of the London penetration depth at five doping levels. The solid curves are calculated theoretically. The symbolic curves are taken from experimental work on La$_{2-x}$Sr$_{x}$CuO$_{4}$ \cite{Panagopoulos_1999}. Matching to one level of doping for solid and symbolic curves is indicated by the same color. The magnitudes of doping $x$ are
indicated next to corresponding symbols. The insert shows doping
dependence of the critical temperature. The model parameters (in eV): $\tau=0.225$, $t=0.12$, $J=1.76$, $I=0.118$, $U_p=3.3$, $V_2=0.1$. $V_1$ is not specified since it does not contribute to d-wave pairing in the SFM. The phase: $\alpha_x=0.002$.}\label{fig-3}
\end{center}
\end{figure}

Figure \ref{fig-3} compares the temperature dependencies of $\lambda^{-2}$ obtained at different doping within the SFM model (solid lines) with that taken from the experiment on La$_{2-x}$Sr$_x$CuO$_4$ \cite{Panagopoulos_1999} (symbolic curves). The $T_c$$-$$x$ phase diagram, shown in the insert, is obtained within the spin-polaron approach and correlates well to the experimental phase diagram for LSCO superconductors in both the left boundary of the superconducting dome at  $x\cong0.05$ and the maximum critical temperature $T_{max}$=39 K. At the same time, the right boundary of the theoretical dome exceeds that of experimental dome by the value of about 0.1. The reason for this is that in the present study, we adopted the low-density approximation, and hence in the strongly overdoped regime our theory seems to be insufficient. As a result a theoretical curve $\lambda^{-2}(T)$ for large doping $x=0.24$ significantly differs from the experimental one since in real LSCO at $x=0.24$ the critical temperature $T_c=20$ K, but according to the phase diagram shown in the insert $T_c=30$ K.

A comparison of the temperature curves of $\lambda^{-2}$ for the same doping $x$ in the figure \ref{fig-3} shows that the values of $T_c$ and $\lambda^{-2}(T=0)$ are on the whole well reproduced for $x=0.15$$-$$0.22$. It can be seen from the figure that all the theoretical temperature dependencies $\lambda^{-2}(T)$, except for $x=0.10$, are slightly convex, as in most experiments on cuprate superconductors \cite{Sonier_1999,Panagopoulos_1999,Khasanov_PRL_2007}. For the doping level $x=0.10$ (the lowest solid curve in the figure \ref{fig-3}) the form of $\lambda^{-2}(T)$ is concave over the entire temperature range what seems to be incompatible with corresponding experimental curve measured in \cite{Panagopoulos_1999}. This discrepancy is most likely due to
the strong spin-charge fluctuations which are well developed in the strongly underdoped regime and which, in particular, result in pseudogap (PG) behavior in cuprates. The present theory is, however, a mean field theory, it does not take into account these spin-charge fluctuations and therefore PG behavior. Since, however, the PG is weak at optimal and higher doping $x\geqslant 0.15$, we are confident that our results for $x=0.15$$-$$0.22$ will be unaffected by the PG behavior.

The comparison of the calculated temperature dependencies $\lambda^{-2}(T)$ on Figure \ref{fig-3} with the corresponding curves from our previous paper \cite{Dzebisashvili_2018} leads to the conclusion that the main effect of taking into account the Coulomb interaction is the decrease of $T_c$. It is important that the main result of \cite{Dzebisashvili_2018}, the inflection point associated with the change of curvature of the function $\lambda^{-2}(T)$ and found experimentally in a number of compounds \cite{Khasanov_PRL_2007,Wojek_2011,Sonier_1999,Khasanov_2007,Anukool_2009,Howald_2018}
remained unaffected. This inflection point was considered as a confirmation of the spin-polaron concept of quasiparticles in cuprate HTSCs.

\section{Conclusion}\label{sec:conclusion}

Within the spin polaron concept, the effect of Coulomb repulsion on modification of the temperature dependence of the London penetration depth $\lambda$ in cuprate high-temperature superconductors was studied.

When obtaining expressions for calculating $\lambda$ two types of Coulomb interactions were taken into account: 1) Hubbard repulsion of two holes on one site and 2) Coulomb repulsion of two holes located on the next-nearest-neighbor oxygen ions. The interaction of the holes on the nearest-neighbor sites was not taken into account because, according to the results of \cite{VDKB_2016}, it does not contribute to the d-wave superconductivity within spin-fermion model.

The calculation of the London penetration depth $\lambda$ was carried out on the basis of the method developed by the authors in \cite{Dzebisashvili_2018} in the framework of the spin-polaron approach, which takes into account the strong coupling between the charge and spin degrees of freedom, as well as the real structure of the CuO$_2$-planes with two oxygen ions per unit cell.

On the basis of numerical calculations of the temperature dependence of inverse square of the London penetration depth, carried out with the generally accepted values of the Emery model parameters, it was shown that taking into account the Coulomb interaction results in, as expected from \cite{VVV_DDM_etal_2017,VDKB_JLTP_2018}, a significant decrease in the critical temperature corresponding to zeros of the function $\lambda^{-2}(T)$. This circumstance enabled one to achieve substantially better agreement of the theoretical curves with experimental results \cite{Panagopoulos_1999}, in rather broad range for $x$ around optimal doping ($x=0.15$, 0.20 and 0.22). At the same time for strongly overdoped and   underdoped compounds our results for $\lambda^{-2}(T)$ reveal discrepancy with experimental data. We argue that for large doping ($x=0.24$) this discrepancy is because in our theory the low-density approximation was adopted and hence for doping as large as $x=0.24$ this approximation may be insufficient. On the other hand, the strong spin-charge fluctuations, which in the low doping regime are well developed  due to proximity to antiferromagnetic region, are not taken into account properly in our theory. We suggest this to be the main reason for discrepancy of our results for $\lambda^{-2}(T)$ with experimental one at doping as small as $x=0.10$.

However, for cuprates with moderate doping $x=0.15$, $0.20$, and $0.22$ the proposed theory describes the experimental dependencies $\lambda^{-2}(T)$ quite well and clearly shows that accounting for the Coulomb interaction leads to an almost three-fold decrease in the value of $T_c$, but does not change the functional form of the temperature dependence of $\lambda^{-2}$, which was obtained earlier. In particular, the inflection point of the function $\lambda^{-2}(T)$, whose existence is considered by us as a confirmation of the spin-polaron nature of the quasiparticles in cuprates, remained intact.
\bigskip

\ack

The work was financially supported by the Russian Foundation for Basic Research (project \#18-02-00837, \#20-32-70059), the Government of Krasnoyarsk Region, the Krasnoyarsk Regional Science and Technology Support Fund (projects: \#18-42-240014 "One-orbit effective model of an ensemble of spin-polaron quasiparticles in the problem of describing the intermediate state and pseudogap behavior of cuprate superconductors" and \#18-42-243002 "Manifestation of spin-nematic correlations in spectral characteristics of electronic structure and their influence on practical properties of cuprate superconductors"). The work of K.K.K. was supported by the Council for Grants of the President of the Russian Federation (project MK-1641.2020.2).

\appendix
\section{} 
\label{sec:Appendix}

The equations for the components of the superconducting order parameter, which are discussed at the beginning of the Section \ref{sec:EqOP}, have the form
\begin{eqnarray} \label{Deltas}
\Delta_{1k}=&-\bigl(\cos k_x-\cos k_y\bigr)\frac{2V_2}{N}\sum_{q}\cos q_x
  \langle a_{q\uparrow}a_{-q\downarrow}\rangle,\nonumber\\
\Delta_{4k}=&-\bigl(\cos k_x-\cos k_y\bigr)\frac{2V_2}{N}\sum_{q}\cos q_x
  \langle b_{q\uparrow}b_{-q\downarrow}\rangle,\nonumber\\
\Delta_{5k}=&+\bigl(\cos k_x-\cos k_y\bigr)\frac{I}{N}\sum_{q}
  \bigl(\cos q_x-\cos q_y\bigr)\nonumber\\
&\qquad\times\bigl(\langle L_{q\uparrow}L_{-q\downarrow}\rangle
  -C_1\langle u_{q\uparrow}u_{-q\downarrow}\rangle\bigr)\nonumber\\
&+\frac{U_p}{N}\sum_q\frac{C_1}{2}\bigl(\cos(k_x-2\alpha_x)
  \langle a_{q\uparrow}a_{-q\downarrow}\rangle\nonumber\\
&\quad\qquad\qquad+\cos k_y\langle b_{q\uparrow}b_{-q\downarrow}\rangle\bigr)\nonumber\\
&-\bigl(\cos k_x-\cos k_y\bigr)\frac{V_2}{N}\sum_q C_1 \cos q_x \nonumber\\
&\qquad\times\bigl(\langle a_{q\uparrow}a_{-q\downarrow}\rangle
  +\langle b_{q\uparrow}b_{-q\downarrow}\rangle\bigr),
\end{eqnarray}
where
\begin{eqnarray}
\langle u_{q\uparrow}u_{-q\downarrow}\rangle=
&-s^2_{q,x}\langle a_{q\uparrow}a_{-q\downarrow}\rangle
 -s^2_{q,y}\langle b_{q\uparrow}b_{-q\downarrow}\rangle\nonumber\\
&-s_{q,x}s_{q,y}\bigl(\langle a_{q\uparrow}b_{-q\downarrow}\rangle
  +\langle b_{q\uparrow}a_{-q\downarrow}\rangle\bigr).
\end{eqnarray}

Functions $S_{ij}^{(l)}(k,\omega)$ used when writing expressions (\ref{eqMnm}) are defined as
\begin{eqnarray}\label{eqsSij}
S^{(1)}_{11}(k,\omega)&=+Q_{3y}(k,-\omega)Q_{3y}(k,\omega),\nonumber\\
S^{(1)}_{21}(k,\omega)&=+S^{(1)}_{12}(k,-\omega)=Q_{3}(k,-\omega)Q_{3y}(k,\omega),\nonumber\\
S^{(4)}_{11}(k,\omega)&=+S^{(1)}_{22}(k, \omega)=Q_{3}(k,-\omega)Q_{3 }(k,\omega),\nonumber\\
S^{(5)}_{11}(k,\omega)&=-Q_{y}(k,-\omega)Q_{y }(k,\omega),\nonumber\\
S^{(4)}_{12}(k,\omega)&=+Q_{3}(k,-\omega)Q_{3x}(k,\omega),\nonumber\\
S^{(4)}_{21}(k,\omega)&=+S^{(4)}_{12}(k,-\omega),\nonumber\\
S^{(5)}_{12}(k,\omega)&=-Q_{y}(k,-\omega)Q_{x}(k,\omega),\nonumber\\
S^{(5)}_{21}(k,\omega)&=+S^{(5)}_{12}(k,-\omega),\nonumber\\
S^{(4)}_{22}(k,\omega)&=+Q_{3x}(k,-\omega)Q_{3x}(k,\omega),\nonumber\\
S^{(5)}_{22}(k,\omega)&=-Q_{x}(k,-\omega)Q_{x}(k,\omega),\nonumber\\
S^{(1)}_{33}(k,\omega)&=-K_k^2S^{(5)}_{11}(k,\omega),\nonumber\\
S^{(4)}_{33}(k,\omega)&=+K_k^2S^{(5)}_{22}(k,\omega),\nonumber\\
S^{(5)}_{33}(k,\omega)&=+Q_{xy}(k,-\omega)Q_{xy}(k,\omega),
\end{eqnarray}
where
\begin{eqnarray}
Q_{x(y)}(k,\omega)&=\bigl(\omega-\xi_{x(y)}\bigr)J_{y(x)}+t_kJ_{x(y)},\nonumber\\
Q_3(k,\omega)&=\bigl(\omega-\xi_L\bigr)t_k+J_xJ_yK_k,\nonumber\\
Q_{3x(3y)}(k,\omega)&=\bigl(\omega-\xi_L\bigr)\bigl(\omega-\xi_{x(y)}\bigr)-J_{x(y)}^2K_k,\nonumber\\
Q_{xy}(k,\omega)&=\bigl(\omega-\xi_x\bigr)\bigl(\omega-\xi_y\bigr)-t_k^2.
\end{eqnarray}

The function $R^x(k,\omega)$, which is included in the equation for the chemical potential (\ref{Eq_mu}), is defined as follows
\begin{eqnarray}\label{eqqRx}
&R^x(k,\omega)=\bigl(Q_{3y}(k,\omega)+Q_{3x}(k,\omega)\bigr)\Psi(k,\omega)\nonumber\\
&\quad-2\bigl(J_xQ_{y}(k,-\omega)\Delta^*_{1k}+J_yQ_x(k,-\omega)\Delta^*_{4k}\bigr)\Delta^*_{5k}\nonumber\\
&\quad-\bigl(\omega-\xi_L\bigr)\bigl(Q_{3y}(k,-\omega){\Delta^*_{1k}}^2+Q_{3x}(k,-\omega){\Delta^*_{4k}}^2\bigr)\nonumber\\
&\quad-\bigl(2\omega-\xi_x-\xi_y\bigr)Q_{xy}(k,-\omega){\Delta^*_{5k}}^2/K_k^2,\\\label{eqPsi}
&\Psi(k,\omega)=\bigl(\omega+E_k\bigr)\bigl(\omega+\epsilon_{2,-k}\bigr)\bigl(\omega+\epsilon_{1,-k}\bigr).
\end{eqnarray}

Thermodynamic averages of equation (\ref{curr_den}) are defined by the expressions
\begin{eqnarray}\label{eqCor_abL}
&\langle a^{\dag}_{k\alpha}a_{k\alpha}\rangle
 =Q_{3y}(k,\omega)\Psi(k,\omega)-2J_yQ_x(k,-\omega)\Delta^*_{4k}\Delta^*_{5k}\nonumber\\
&\quad-\bigl(\omega-\xi_L\bigr)Q_{3x}(k,-\omega){\Delta^*_{4k}}^2\nonumber\\
&\quad-\bigl(\omega-\xi_y\bigr)Q_{xy}(k,-\omega){\Delta^*_{5k}}^2/K_k^2,\nonumber\\
&\langle a^{\dag}_{k\alpha}b_{k\alpha}\rangle
 =Q_{3}(k,\omega)\Psi(k,\omega)
 +J_xQ_{x}(k,-\omega)\Delta^*_{4k}\Delta^*_{5k}\nonumber\\
&\quad+J_yQ_{y}(k,-\omega)\Delta^*_{1k}\Delta^*_{5k}
 -t_kQ_{xy}(k,-\omega){\Delta^*_{5k}}^2/K_k^2\nonumber\\
&\quad+\bigl(\omega-\xi_L\bigr)Q_{3}(k,-\omega)\Delta^*_{1k}\Delta^*_{4k},\nonumber\\
&\langle a^{\dag}_{k\alpha}L_{k\alpha}\rangle
 =Q_y(k,\omega)K_k\Psi(k,\omega)
 +t_kQ_x(k,-\omega)\Delta^*_{4k}\Delta^*_{5k}\nonumber\\
&\quad+J_yQ_3(k,-\omega)K_k\Delta^*_{1k}\Delta^*_{4k}
-J_xQ_{3x}(k,-\omega)K_k{\Delta^*_{4k}}^2\nonumber\\
&\quad+\bigl(\omega-\xi_y\bigr)Q_y(k,-\omega)\Delta^*_{1k}\Delta^*_{5k},
\end{eqnarray}
where $\Psi(k,\omega)$ is defined in (\ref{eqPsi}).


\section*{References}

\begin{thebibliography}{55}
  \bibitem{Emery_1987}\label{bib:Emery_1987}
    {Emery V J}
    1987
    {\it Phys. Rev. Lett.}
    {\bf 58} 2794
  \bibitem{Varma_1987}\label{bib:Varma_1987}
    {Varma C M,  Schmitt-Rink S and  Abrahams E}
    1987
    {\it Solid State Commun.}
    {\bf 62} 681
  \bibitem{Hirsch_1987}\label{bib:Hirsch_1987}
    {Hirsch J E}
    1987
    {\it Phys. Rev. Lett.}
    {\bf 59} 228
  \bibitem{Gaididei_1988}\label{bib:Gaididei_1988}
    {Gaididei Yu B and Loktev V M}
    1988
    {\it Phys. Status Solidi B}
    {\bf 147} 307
  \bibitem{Ovchinnikov_1989}\label{bib:Ovchinnikov_1989}
    {Ovchinnikov S G and Sandalov I S}
    1989
    {\it Physica C}
    {\bf 161} 607
  \bibitem{Barabanov_1988}\label{bib:Barabanov_1988}
    {Barabanov A F, Maksimov L A and Uimin G V}
    1988
    {\it JETP Lett.}
    {\bf 47} 622
  \bibitem{Zaanen_1988}\label{bib:Zaanen_1988}
    {Zaanen J and Ole\'{s} A M}
    1988
    {\it Phys. Rev. B}
    {\bf 37} 9423
  \bibitem{Emery_Reiter_1988}\label{bib:Emery_Reiter_1988}
    {Emery V J and Reiter G}
    1988
    {\it Phys. Rev. B}
    {\bf 38} 4547
  \bibitem{Prelovsek_1988}\label{bib:Prelovsek_1988}
    {Prelov\v{s}ek P}
    1988
    {\it Phys. Lett. A}
    {\bf 126} 287
  \bibitem{Stechel_1988}\label{bib:Stechel_1988}
    {Stechel E B and Jennison D R}
    1988
    {\it Phys. Rev. B}
    {\bf 38} 4632
  \bibitem{Kohno_2018}\label{bib:Kohno_2018}
    {Kohno M}
    2018
    {\it Rep. Prog. Phys.}
    {\bf 81} 042501
  \bibitem{Kitatani_2019}\label{bib:Kitatani_2019}
    {Kitatani M, Schafer T, Aoki H and Held K}
    2019
    {\it Phys. Rev. B}
    {\bf 99} 041115(R)
  \bibitem{Spalek_2017}\label{bib:Spalek_2017}
    {Spalek J, Zegrodnik M and Kaczmarczyk J}
    2017
    {\it Phys. Rev. B}
    {\bf 95} 024506
  \bibitem{Barabanov_1993}\label{bib:Barabanov_1993}
    {Barabanov A F, Maksimov L A and Zhukov L E}
    1993
    {\it Physica C}
    {\bf 212} 375
  \bibitem{Barabanov_1996}\label{bib:Barabanov_1996}
    {Barabanov A F, Berezovskii V M, Zasinas E and Maksimov L A}
    1996
    {\it JETP}
    {\bf 83} 819
  \bibitem{Barabanov_1997}\label{bib:Barabanov_1997}
    {Barabanov A F, Kuzian R O and Maksimov L A}
    1997
    {\it Phys. Rev. B}
    {\bf 55} 4015
  \bibitem{Starykh_1995}\label{bib:Starykh_1995}
    {Starykh O A, de Alcantara Bonfim O F and Reiter G F}
    1995
    {\it Phys. Rev. B}
    {\bf 52} 12534
  \bibitem{Barabanov_Qpol_1997}\label{bib:Barabanov_Qpol_1997}
    {Barabanov A F, Zasinas E, Urazaev OV and Maksimov L A}
    1997
    {\it JETP Lett.}
    {\bf 66} 182
  \bibitem{Barabanov_2001}\label{bib:Barabanov_2001}
    {Barabanov A F, Kovalev A A, Urazaev O V, Belemuk A M and Hayn R}
    2001
    {\it JETP}
    {\bf 92} 677
  \bibitem{Kuzian_2003}\label{bib:Kuzian_2003}
    {Kuzian R O, Hayn R and Barabanov A F}
    2003
    {\it Phys. Rev. B}
    {\bf 68} 195106
  \bibitem{DVB_2013}\label{bib:DVB_2013}
    {Dzebisashvili D M, Val'kov V V and Barabanov A F}
    2013
    {\it JETP Lett.}
    {\bf 98} 528
  \bibitem{VDB_PLA_2015}\label{bib:VDB_PLA_2015}
    {Val'kov V V, Dzebisashvili D M and Barabanov A F}
    2015
    {\it Phys. Lett. A}
    {\bf 379} 421
  \bibitem{VDB_JLTP_2015}\label{bib:VDB_JLTP_2015}
    {Val'kov V V, Dzebisashvili D M and Barabanov A F}
    2015
    {\it J. Low Temp. Phys.}
    {\bf 181} 134
  \bibitem{VDB_JSNM_2016}\label{bib:VDB_JSNM_2016}
    {Val'kov V V, Dzebisashvili D M and Barabanov A F}
    2016
    {\it J. Supercond. Nov. Magn.}
    {\bf 29} 1049
  \bibitem{Dzebisashvili_2018}\label{bib:Dzebisashvili_2018}
    {Dzebisashvili D M and Komarov K K}
    2018
    {\it Eur. Phys. J. B}
    {\bf 91} 278
  \bibitem{Khasanov_PRL_2007}\label{bib:Khasanov_PRL_2007}
    {Khasanov R, Shengelaya A, Maisuradze A, Mattina F La, Bussmann-Holder A, Keller H and Muller K A}
    2007
    {\it Phys. Rev. Lett.}
    {\bf 98} 057007
  \bibitem{Wojek_2011}\label{bib:Wojek_2011}
    {Wojek B M, Weyeneth S, Bosma S, Pomjakushina E and Puzniak R}
    2011
    {\it Phys. Rev. B}
    {\bf 84} 144521
  \bibitem{Sonier_1999}\label{bib:Sonier_1999}
    {Sonier J E, Brewer J H, Kiefl R F, Morris G D, Miller R I, Bonn D A, Chakhalian J, Heffner R H, Hardy W N and Liang R}
    1999
    {\it Phys. Rev. Lett.}
    {\bf 83} 4156
  \bibitem{Khasanov_2007}\label{bib:Khasanov_2007}
    {Khasanov R, Strassle S, Castro D Di, Masui T, Miyasaka S, Tajima S, Bussmann-Holder A and Keller H}
    2007
    {\it Phys. Rev. Lett.}
    {\bf 99} 237601
  \bibitem{Anukool_2009}\label{bib:Anukool_2009}
    {Anukool W, Barakat S, Panagopoulos C and Cooper J R}
    2009
    {\it Phys. Rev. B}
    {\bf 80} 024516
  \bibitem{Panagopoulos_1999}\label{bib:Panagopoulos_1999}
    {Panagopoulos C, Rainford B D, Cooper J R, Lo W, Tallon J L, Loram J W, Betouras J, Wang Y S and Chu C W}
    1999
    {\it Phys. Rev. B}
    {\bf 60} 14617
  \bibitem{VVV_DDM_etal_2017}\label{bib:VVV_DDM_etal_2017}
    {Val'kov V V, Dzebisashvili D M, Korovushkin M M and Barabanov A F}
    2017
    {\it JETP}
    {\bf 125} 810
  \bibitem{Hybertsen_1989}\label{bib:Hybertsen_1989}
    {Hybertsen M S, Schluter M and Christensen N E}
    1989
    {\it Phys. Rev. B}
    {\bf 39} 9028
  \bibitem{Ogata_2008}\label{bib:Ogata_2008}
    {Ogata M and Fukuyama H}
    2008
    {\it Rep. Prog. Phys.}
    {\bf 71} 036501
  \bibitem{Fischer_2011}\label{bib:Fischer_2011}
    {Fischer M H and Kim E-A}
    2011
    {\it Phys. Rev. B}
    {\bf 84} 144502
  \bibitem{VDKB_JMMM_2017}\label{bib:VDKB_JMMM_2017}
    {Val'kov V V, Dzebisashvili D M, Korovushkin M M and Barabanov A F}
    2017
    {\it J. Magn. Magn. Mat.}
    {\bf 440} 123
  \bibitem{Barabanov_2011}\label{bib:Barabanov_2011}
    {Barabanov A F, Mikheenkov A V and Shvartsberg A V}
    2011
    {\it Theor. Math. Phys.}
    {\bf 168} 1192
  \bibitem{Shimahara_1991}\label{bib:Shimahara_1991}
    {Shimahara H and Takada S}
    1991
    {\it JPSJ}
    {\bf 60} 2394
  \bibitem{Peierls_1933}\label{bib:Peierls_1933}
    {Peierls R}
    1933
    {\it Z. Physik}
    {\bf 80} 763
  \bibitem{Lifshitz_1978}\label{bib:Lifshitz_1978}
    {Lifshitz E M and Pitaevskii L P}
    1980
    {\it Statistical Physics: Theory of the Condensed State (Part 2)}
    {(Oxford: Pergamon Press plc)}
    {p 387}
  \bibitem{Schriffer_1964}\label{bib:Schriffer_1964}
    {Schriffer J R}
    1964
    {\it Theory of Superconductivity}
    {(New York: Benjamin)}
    {p 332}
  \bibitem{Tinkham_1996}\label{bib:Tinkham_1996}
    {Tinkham M}
    1996
    {\it Introduction to Superconductivity}
    {(New York: McGraw-Hill)}
    {p 454}
  \bibitem{Zwanzig_1961}\label{bib:Zwanzig_1961}
    {Zwanzig R}
    1961
    {\it Phys. Rev.}
    {\bf 124} 983
  \bibitem{Mori_1965}\label{bib:Mori_1965}
    {Mori H}
    1965
    {\it Prog. Theor. Phys.}
    {\bf 33} 423
  \bibitem{Roth_1968}\label{bib:Roth_1968}
    {Roth L M}
    1968
    {\it Phys. Rev. Lett.}
    {\bf 20} 1431
  \bibitem{Roth_1969}\label{bib:Roth_1969}
    {Roth L M}
    1969
    {\it Phys. Rev.}
    {\bf 184} 451
  \bibitem{Rowe_1968}\label{bib:Rowe_1968}
    {Rowe D J}
    1968
    {\it Rev. Mod. Phys.}
    {\bf 40} 153
  \bibitem{Tserkovnikov_1981}\label{bib:Tserkovnikov_1981}
    {Tserkovnikov Y A}
    1981
    {\it Teor. Mat. Fiz.}
    {\bf 49} 219;
    {Tserkovnikov Y A}
    1982
    {\it Theoretical and Mathamatical Physics}
    {\bf 52} 712
  \bibitem{Plakida_book_2010}\label{bib:Plakida_book_2010}
    {Plakida N M}
    2010
    {\it High-Temperature Cuprate Superconductors}
    {(Berlin: Springer)}
    {p 570}
  \bibitem{Mancini_2004}\label{bib:Mancini_2004}
    {Mancini F and Avella A}
    2004
    {\it Adv. Phys.}
    {\bf 53} 537
  \bibitem{Izyumov_1999}\label{bib:Izyumov_1999}
    {Izyumov Yu A}
    1999
    {\it Phys. Usp.}
    {\bf 42} 215
  \bibitem{VDKB_2016}\label{bib:VDKB_2016}
    {Val'kov V V, Dzebisashvili D M, Korovushkin M M and Barabanov A F}
    2016
    {\it JETP Lett.}
    {\bf 103} 385
  \bibitem{Zubarev_1960}\label{bib:Zubarev_1960}
    {Zubarev D N}
    1960
    {\it Sov. Phys. Usp.}
    {\bf 3} 320
  \bibitem{VDKKB_2019}\label{bib:VDKKB_2019}
    {Val'kov V V, Dzebisashvili D M, Korovushkin M M, Komarov K K and Barabanov A F}
    2019
    {\it JETP}
    {\bf 128} 885
  \bibitem{VDKB_JLTP_2018}\label{bib:VDKB_JLTP_2018}
    {Val'kov V V, Dzebisashvili D M, Korovushkin M M and Barabanov A F}
    2018
    {\it J. Low Temp. Phys.}
    {\bf 191} 408
  \bibitem{Howald_2018}\label{bib:Howald_2018}
    {Howald L, Stilp E, Baiutti F and Dietl C \etal}
    2018
    {\it Phys. Rev. B}
    {\bf 97} 094514
\end{thebibliography}
\end{document}